\documentclass[11pt]{article}
\usepackage{multicol}
\usepackage{amsmath,amsfonts,amssymb,amsthm,array}
\usepackage{subcaption}
\usepackage[margin=1in]{geometry}
\usepackage[colorlinks=true,allcolors=blue]{hyperref}
\usepackage{bbold}
\usepackage{subcaption}
\usepackage{titlesec}
\usepackage[toc,page]{appendix}
\usepackage{bbm}
\usepackage[toc,page]{appendix}
\usepackage{xspace}
\usepackage{graphicx}
\usepackage{amsmath}
\usepackage{hyperref}
\usepackage{braket}
\usepackage{authblk}
\usepackage{appendix}
\usepackage{hyperref}
\usepackage{xcolor}
\title{Robustness of deformed catlike states under dissipative decoherence}
\date{}

\author{Abdessamad Belfakir\thanks{abdobelfakir01@gmail.com}}
\author{\hspace{0.5cm}Adil Belhaj\thanks{belhajadil@fsr.ac.ma}}
\author{\hspace{0.2cm}Yassine Hassouni\thanks{yassine.hassouni@gmail.com}}
\affil{Equipe des Sciences de la Mati\`{e}re et du Rayonnement (ESMaR), \\Facult\'e des Sciences, Universit\'e  Mohammed V de Rabat,
Av. Ibn Battouta, B.P.1014, Agdal, Rabat, Morocco.}

\begin{document}
\maketitle
\begin{abstract}
The generation of coherent superposition of distinct physical systems and the construction
of robust entangled states under decoherence are the most experimental challenges of quantum technologies. In this work, we investigate the behaviors of catlike states of a deformed harmonic oscillator under dissipative decoherence. Varying the deformation parameters, we obtain catlike states having more resistance against decoherence than catlike states of
the ordinary harmonic oscillator. Furthermore, we study nonclassical properties and entanglement of different catlike states subjects to decoherence caused by a dissipative interaction with a large environment. Depending
on different deformation  parameters, we reveal that the nonclassical properties
of catlike states under dissipative interaction can be more retarded and preserved
in the time.
\end{abstract}
{\quotation\noindent{\bf Keywords:}
Catlike states, decoherence, nonclassical properties, entanglement.

\endquotation}

	\section{Introduction }\label{introduction}
Coherent states or quasi-classical states of the harmonic oscillator were first introduced  by Schr\"odinger in order to make a connection between the classical and the  quantum formalisms \cite{shrodinger}. It has been shown that these states not only minimize the Heisenberg uncertainty inequality for position and momentum operators but also maintain maximum localizability during the time evolution of the harmonic oscillator \cite{Gazeau,Zhang}. In particular, their dispersions  on  the kinetic energy and on the  potential energy are identical \cite{Gazeau}. The importance and the physical applicability of these states, in quantum optics, have been investigated in many works including the papers of Glauber, Klauder  and Sudarshan \cite{Glauber,Klauder1,Klauder2,PhysRevLett.10.277}. The concept of coherent states has not been restricted to only the harmonic oscillator and has been implemented to other physical systems 	\cite{PhysRevA.54.4560,Klauder3,Hassouni2}. In this way, these states are called nonlinear coherent states and have been constructed for certain physical systems \cite{Hassouni2,Curado2,hydrogenatom}. Furthermore, it  has been shown that coherent states can be built for any Lie symmetry \cite{Perelomov,Klau,Gilmor}. Particularly, the su(1,1) coherent states related to the SU(1,1) group and the su(2) coherent states have been constructed and applied in quantum optics  and in condensed matter physics \cite{Zhang,Perelomov,Inomata}. It is worth noting that there are two different approaches to elaborate su(1,1) coherent states. The first one is called Perelomov approach which is based on the application of the displacement operator on the ground state \cite{Perelomov}. The second one is associated with  Barut-Girardello coherent states  defined as eigenstates of the  annihilation operator of the su(1,1) algebra \cite{Barut}. Several mathematical generalizations of coherent states have been introduced \cite{Gazeau,Hassouni2,Shi,Angelova4,Roy,Maia,Daoud,PhysRevA.64.013817,Madouri,Jurco}. In particular, the generalized Heisenberg algebra (GHA) nonlinear coherent states  have been constructed \cite{Hassouni2} and  studied for several physical systems including the square well potential \cite{Hassouni2}, the P\"oschl-Teller potential \cite{Curado2} and Morse potential \cite{Angelova4}.\\
Recently, a  considerable attention is paid to the experimental observations of nonclassical properties of quantum systems such as squeezing \cite{Slusher}, photon anti-bunching \cite{Kimble,Mandel} and entanglement \cite{Yu}. The latter is recognized as a key resource of quantum computing, quantum cryptography and quantum communications \cite{Nielsen,Horo}. In contrast to classical systems whose states are always a mixture, quantum systems can be prepared in a superposition of quantum states. This property is an interesting quantum effect being very hard to be prepared and  detected due to the decoherence  occurring when the quantum system interacts with a relevant environment \cite{Gardiner}. This interaction causes the loss of  coherence of the quantum system  generating classical mixture states. This mapping is called the decoherence \cite{Gardiner}.  The process of decoherence has been studied both theoretically and experimentally \cite{Neumann,Nemes,Zurek,Retamal,Kampen,Palma,Leonid} for several physical systems such as the trapped ion in many works \cite{Mancini,Filho,Vogel,Meekhof,King}.\\ The main aim of this work is to contribute to these activities by investigating the generalized su(1,1) coherent states superposition for a four-parameter perturbed oscillator \cite{2013,Twareque}, known as catlike states or even and odd coherent states. Then, we study their nonclassical properties under decoherence in order to compare them with those of the generalized Heisenberg algebra catlike states. In particular, we discuss the robustness of generalized catlike states against  decoherence caused by the interaction with an environment associated with an infinite collection of harmonic oscillators by using the fidelity. Concretely, we show that the resistance of generalized catlike states depends on the corresponding algebraic structure. Furthermore, we study the entanglement degree of generalized catlike states  under decoherence.\\
This paper is organized as follows.  In section (\ref{su(1,1) deformed states superposition}),  we briefly recall the GHA and the generalized su(1,1) algebra and construct the generalized  su(1,1) catlike states for a perturbed harmonic oscillator. Furthermore, in section (\ref{sec4}) we study the resistance against decoherence of the constructed catlike states with that of the GHA catlike states. Moreover, in section (\ref{sec5}) we study  the effect of the decoherence on  photon distribution function, statistical distribution and quantum entanglement of generalized catlike states in terms of various parameters of deformation. Finally, our conclusions are given in section (\ref{sec6}).
	\section{GHA, generalized su(1,1) algebra and coherent states superposition}\label{su(1,1) deformed states superposition}
		\subsection{Generalities on GHA}
To start, we give a concise review on the GHA by presenting its essential aspects \cite{Monteiro1,Monteiro2,Hassouni,Abdessamad1}. It is recalled that the GHA is described by the generators $H$, $A^\dagger$ and $A$ satisfying
	\begin{equation}\label{A+}
HA^\dagger=A^\dagger f(H),\quad AH=f(H)A,
\end{equation}
and
\begin{equation}\label{H}
[A,A^\dagger]=f(H)-H,
\end{equation}
where $H=H^\dagger$ is the dimensionless  Hamiltonian of the physical system under consideration. $(A^\dagger)^\dagger=A$. $f$ is an analytical function of $H$, called the characteristic function of the GHA. The associated Casimir operator reads as
\begin{equation}\label{Cas}
C=A^\dagger A-H=AA^\dagger-f(H).
\end{equation}
The irreducible representation of the GHA is given through an eigenvector $\ket{n}$ of the Hamiltonian $H$,
\begin{equation}
H\ket{n}=\varepsilon_n\ket{n}\quad\text{such that}\quad\varepsilon_{n+1}=f(\varepsilon_n), \quad \text{for}\quad n=0,1,\dots.
\end{equation}
 The Fock space representation of GHA generators is then given in terms of the lowest energy eigenvalue $\varepsilon_0$. The eigenvalue $\varepsilon_n=f^n(\varepsilon_{0})$ is just the $n$-iterate of $\varepsilon_0$ under the function $f$ \cite{Monteiro1}. By using (\ref{A+})-(\ref{Cas}), one can show that
\begin{equation}\label{A+Action}
A^\dagger\left\vert n\right\rangle=N_{n}\left\vert n+1 \right\rangle,
\end{equation}
\begin{equation}\label{A-Action}
A\left\vert n\right\rangle=N_{n-1}\left\vert n-1 \right\rangle,
\end{equation}
where
\begin{equation}\label{Nnn}
N_{n}^2=\varepsilon_{n+1}-\varepsilon_{0}=f(\varepsilon_{n})-\varepsilon_{0} .
\end{equation}
The operators $A^\dagger$ and $A$ are interpreted as the creation and annihilation operators of GHA, respectively. It is noted that the vacuum state condition $A\ket{0}=0$ is verified. Taking $f(H)=H+1$, where $H$ is the dimensionless Hamiltonian of the harmonic oscillator, the GHA reduces to the ordinary Heisenberg algebra spanned by $H$ and the bosonic ladder operators $a$ and $a^+$ \cite{Monteiro1}. Similarly, considering $f(H)=qH+1$, the relations (\ref{A+})-(\ref{H}) recover the $q$-harmonic oscillator algebra \cite{Monteiro1}. Furthermore, it has been shown that the GHA can be applied for physical systems involving known spectrum such as the square well potential \cite{Hassouni}, the Morse potential \cite{Abdessamad1,Angelova1} and the P\"oschl-Teller potential \cite{Curado2,Bagarello1}.
\subsection{Generalized su(1,1) algebra}
The su(1,1) algebra is a primordial structure in physics since it appears in many formalisms \cite{Wybourne,Lie Groups}. Furthermore, the $q$-deformed su(1,1) algebra has been also constructed and applied widely in many areas of physics \cite{Chakrabarti,Coon,Perelomov2}. Moreover, a generalization of the su(1,1) algebra can be constructed by following the same ideas developed for the GHA and the generalized su(2) introduced  in \cite{Curado00,Curado01}. The generalized su(1,1) algebra is defined by the Hamiltonian $H$ and the ladder operators $J_+$ and $J_-$ satisfying
	\begin{equation}\label{su11+}
HJ_+=J_+ f(H),\quad J_-H=f(H)J_- ,
\end{equation}
\begin{equation}\label{su11H}
[J_+,J_-]=(H-f(H))(H+f(H)-1),
\end{equation}
where $J_+^\dagger=J_-$ and $f(H)$ is a given function of the Hamiltonian $H$. By considering $f(H)=H+1$, this algebra  becomes the standard su(1,1) algebra. The Casimir operator  has now the following form
\begin{equation}\label{cas}
\Gamma=J_+J_--H(H-1)=J_-J_+-f(H)(f(H)-1).
\end{equation}
The representations of the generalized su(1,1) algebra generators can be easily deduced. Let $\varepsilon_{n+1}=f(\varepsilon_n)$. Assuming that $J_-\ket{0}=0$, we can show that
\begin{equation}\label{j++}
J_+\ket{n}=\mathcal{N}_{n}\ket{n+1},
\end{equation}
\begin{equation}\label{j--}
J_-\ket{n}=\mathcal{N}_{n-1}\ket{n-1},
\end{equation}
where
\begin{equation}\label{NSU11}
\mathcal{N}_{n}^2=(\varepsilon_{n+1}-\varepsilon_{0})(\varepsilon_{n+1}+\varepsilon_{0}-1)\quad\text{for}\quad n=0,1,\dots.
\end{equation}
Similarly to the GHA, the generalized su(1,1) algebra involves  the creation and  annihilation operators $J_+$ and $J_-$, respectively.
	\subsection{Construction of generalized su(1,1) coherent states superposition}
Here, we would like to construct the catlike states associated with  the generalized su(1,1) algebra. For this purpose, we first construct the generalized su(1,1) nonlinear coherent state. We define this state as an eigenstate of the annihilation operator $J_-$
	\begin{equation}
J_-\ket{z}=z\ket{z}.
	\end{equation}
	Since  $J_-$ is a non hermitian operator, $z$ is, in general, a complex number. The state $\ket{z}$ is expanded in terms of energy eigenvectors $\ket{n}$ as
\begin{equation}\label{Cstat}
\ket{z}=\sum_{n=0}^\infty c_n\ket{n},
\end{equation}
where $c_n$ are complex coefficients satisfying $\sum_{n=0}^\infty|c_n|^2=1$. The action of $J_-$ on such state, leads to $c_n=\dfrac{c_0z^n}{\mathcal{N}_{n-1}!}$ where $\mathcal{N}_{n}!=\mathcal{N}_n\dots \mathcal{N}_0$ and $\mathcal{N}_{-1}!:=1$ . Let $c_0=N(|z|)$, the state $\ket{z}$ can be rewritten as
	\begin{equation}\label{su11cs}
	\ket{z}=N(|z|)\sum_{n=0}^\infty\dfrac{z^n}{\mathcal{N}_{n-1}!}\ket{n}.
	\end{equation}
	Imposing the normalization condition $\braket{z|z}=1$, we get
		\begin{equation}\label{yop}
N(|z|)=\left(\sum_{n=0}^\infty\dfrac{|z|^{2n}}{(\mathcal{N}_{n-1}!)^2}\right)^{-1/2}.
	\end{equation}
The generalized su(1,1) catlike states can be easily obtained. They are given by
\begin{equation}\label{psi}
\ket{\psi_{\pm}}=\mathcal{N}_{\pm}(|z|)(\ket{z}\pm\ket{-z}),
\end{equation}
where

\begin{equation}\label{SU11CATN}
\mathcal{N}_\pm(|z|)=\left(2\pm 2(N(|z|))^2\sum_{n=0}^\infty\dfrac{(-1)^n|z|^{2n}}{(\mathcal{N}_{n-1}!)^2}\right)^{-1/2}.
\end{equation}
\subsection{Construction of generalized su(1,1) catlike states for a deformed harmonic oscillator}\label{sec3}
As an application of the generalized su(1,1) algebra, we consider  a perturbed harmonic oscillator having the dimensionless energy spectrum  given by
\begin{equation}\label{varepsilon}
\varepsilon_{n}=n+g(n).
\end{equation}
The function $g(n)$ reads as
\begin{equation}
g(n)=\dfrac{an+e}{cn+d},
\end{equation}
where $a,b,c,d$ are real parameters different from zero \cite{2013}. It is a perturbed function associated with the following conditions
\begin{equation}
|a/c|<1,\quad -\dfrac{4ad-4ce}{c^2}\geq r-1 \quad\text{and}\quad \dfrac{d}{c}>0,
\end{equation}
where $r\in[0,1]$. Subject to such conditions, the spectrum (\ref{varepsilon}) is strictly increasing. In this way,  the associated generalized su(1,1) algebra can be constructed. For the simplicity reasons, we take  $c=1$.\\
Taking $n=\varepsilon_{n}+\gamma(\varepsilon_{n})$, the relation (\ref{varepsilon}) gives
\begin{equation}
\varepsilon_{n+1}=n+1+g(n+1)=\varepsilon_{n}+1+\delta(\varepsilon_{n}),
\end{equation}
where
\begin{equation}
\delta(\varepsilon_{n})=\gamma(\varepsilon_{n})+g(\varepsilon_{n}+1+\gamma(\varepsilon_{n})),
\end{equation}
implying that
\begin{equation}
f(\varepsilon_n)=\varepsilon_n+1+\delta(\varepsilon_n).
\end{equation}
Thus, the characteristic function of the generalized su(1,1) algebra is  given by
\begin{equation}\label{fdeformed}
f(H)=H+1+\delta(H).
\end{equation}
The generalized su(1,1) corresponding to the deformed oscillator can be easily obtained by substituting the function (\ref{fdeformed}) in (\ref{su11+})-(\ref{su11H}). It is given as follows
\begin{equation}
[H,J_+]=J_+(1+\delta(H)),
\end{equation}
\begin{equation}
[J_-,J_+]=(1+\delta(H))(2H+\delta(H)).
\end{equation}
By substituting (\ref{varepsilon}) in (\ref{NSU11}), we find that
\begin{equation}\label{Nn-1}
\mathcal{N}_{n-1}!=\left[(\Gamma(d))^2\hspace{0.1cm}d (a d+2 d e+e)\dfrac{n!\Gamma(n+d+a-e/d+1)\Gamma(\alpha+n)\Gamma(\beta+n)}{\Gamma(\alpha+1)\Gamma(\beta+1)\Gamma(d+a-e/d+1)(\Gamma(1+d+n))^2}\right]^{1/2},
\end{equation}
where
\begin{equation}
\alpha=\frac{a+d+1+e/d}{2}-\frac{\sqrt{\left(a d+d^2+d+e\right)^2-4 d (a d+2 d e+e)}}{2 d},
\end{equation}
and
\begin{equation}
\beta=\frac{a+d+1+e/d}{2}+\frac{\sqrt{\left(a d+d^2+d+e\right)^2-4 d (a d+2 d e+e)}}{2 d}.
\end{equation}
The normalization factor given in (\ref{yop}) can be easily computed. It can be  written as
\begin{equation}\label{Factor}N(|z|)=\left[ _2F_3\left(d+1,d+1;\alpha,\beta,a-\frac{e}{d}+d+1;|z|^2\right)\right]^{-1/2},
\end{equation}
where $_2F_3\left(d+1,d+1;\alpha,\beta,a-\frac{e}{d}+d+1;|z|^2\right)$ is the generalized hypergeometric function.
By using (\ref{Nn-1}) and (\ref{Factor}), the normalization factor (\ref{SU11CATN}) associated with the generalized su(1,1) catlike states can be given by
\begin{equation}\label{Fact2}
\mathcal{N}_\pm(|z|)=\left[2\pm2\dfrac{_2F_3\left(d+1,d+1;\alpha,\beta,a-\frac{e}{d}+d+1;-|z|^2\right)}{_2F_3\left(d+1,d+1;\alpha,\beta,a-\frac{e}{d}+d+1;|z|^2\right)}\right]^{-1/2}.
\end{equation}
By substituting (\ref{Nn-1}) and (\ref{Fact2}) in (\ref{psi}), we can construct  the generalized su(1,1) catlike states for the deformed harmonic oscillator. They are given by
\begin{equation}\label{psi1}
\ket{\psi_{\pm}}=\mathcal{N}_{\pm}(|z|)(\ket{z}\pm\ket{-z}).
\end{equation}
It is worth mentioning that the GHA catlike states can be obtained by replacing $\mathcal{N}_{n}$ by $N_n$ in $\ket{z}$ and in $\mathcal{N}_{\pm}(|z|)$.
 In the following, we only consider the catlike states having the form
 \begin{equation}\label{psi2}
 \ket{\psi_{+}}=\mathcal{N}_{+}(|z|)(\ket{z}+\ket{-z}).
 \end{equation}We will see that the  construction of the generalized su(1,1) catlike states for physical systems is advantageous because it gives us the possibility to compare their nonclassical properties under decoherence with the  catlike states associated with the GHA and select the ones whose properties are more preserved and retarded in the time. We will examine particularly  the deformed harmonic oscillator having the spectrum of the form (\ref{varepsilon}). It is noted that the deformed harmonic oscillator becomes the ordinary harmonic oscillator by considering $a=\dfrac{1}{2}$ and $e=\dfrac{d}{2}$. In this case, the GHA catlike states become the ordinary catlike states given by
\begin{equation}\label{psiwal}
\ket{\psi_{\pm}}=(2\pm 2 e^{-2|z|^2})^{-1/2}(\ket{z}\pm\ket{-z}),
\end{equation}
where $\ket{z}$ is now the ordinary Glauber coherent state given by
\begin{equation}
\ket{z}=e^{-|z|^2/2}\sum_{n=0}^\infty\dfrac{z^n}{\sqrt{n!}}\ket{n}.
\end{equation}
\section{Dissipative decoherence of generalized su(1,1) deformed catlike states}\label{sec4}
\begin{figure}
\begin{subfigure}{.5\textwidth}
  \centering
  \includegraphics[width=.8\linewidth]{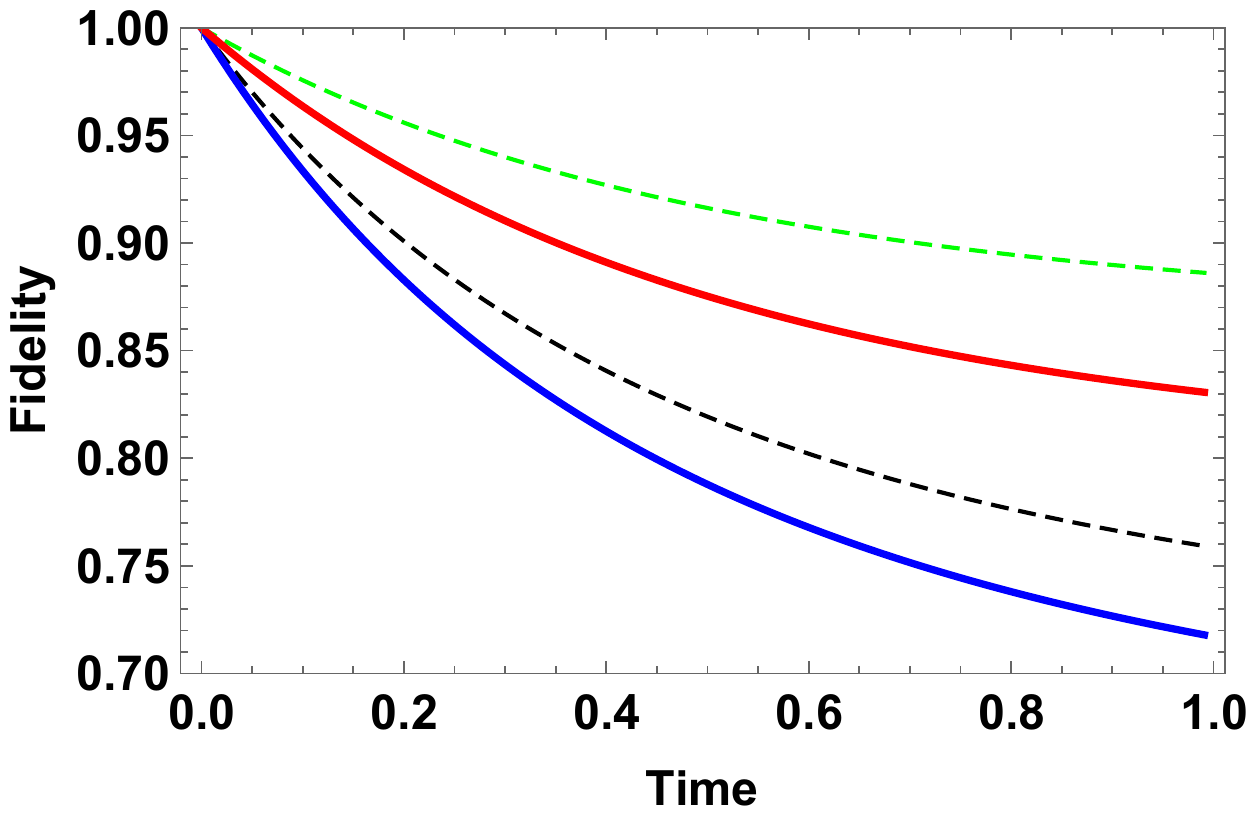}
  \caption{}
  \label{fig1_1}
\end{subfigure}
\begin{subfigure}{.5\textwidth}
  \centering
  \includegraphics[width=.8\linewidth]{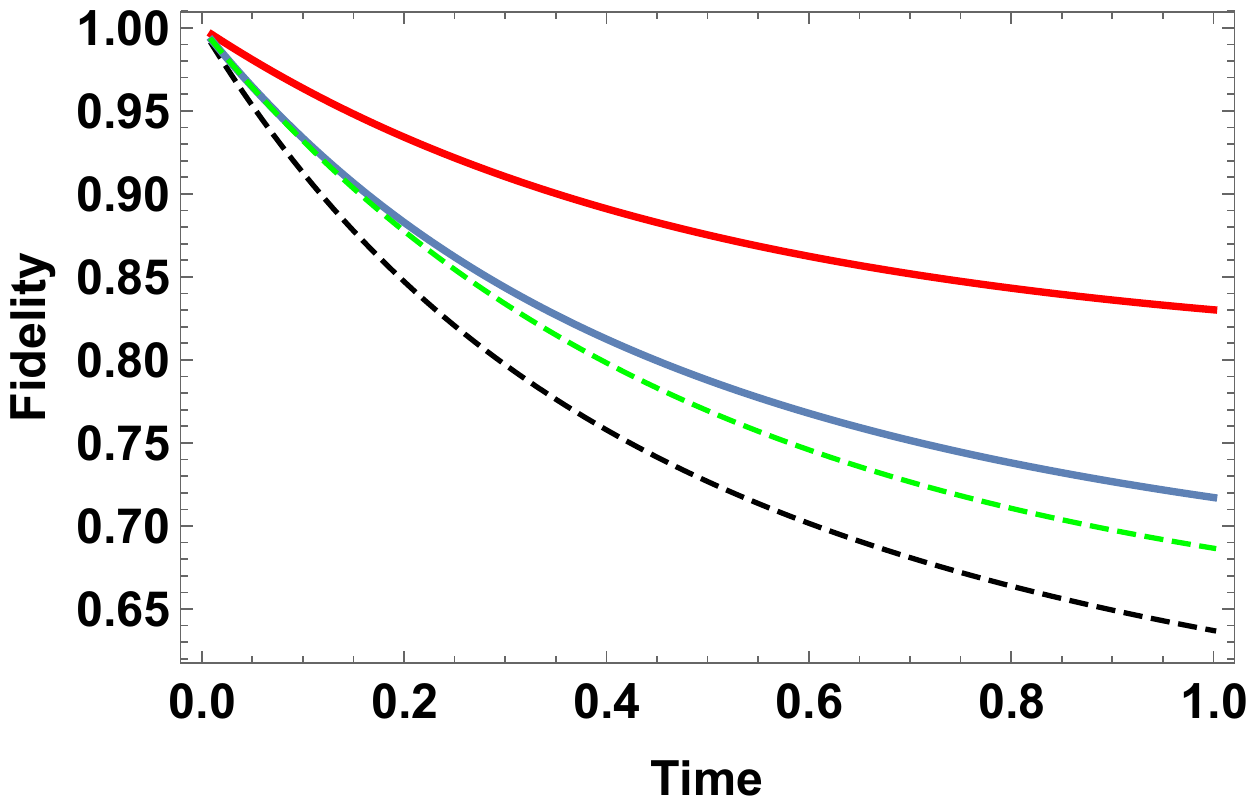}
  \caption{}
   \label{fig1_2}
\end{subfigure}
\\
\begin{subfigure}{.5\textwidth}
  \centering
  \includegraphics[width=.8\linewidth]{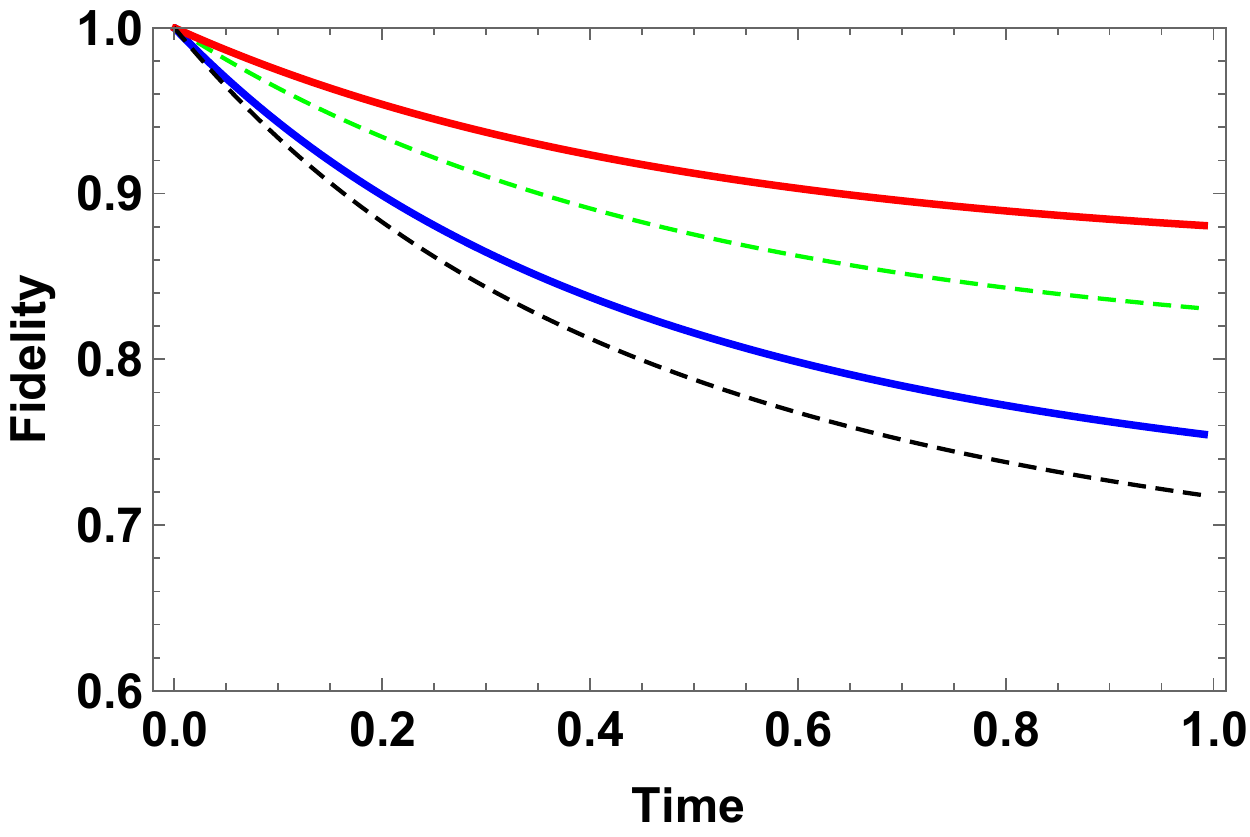}
  \caption{}
   \label{fig1_3}
\end{subfigure}
\begin{subfigure}{.5\textwidth}
  \centering
  \includegraphics[width=.8\linewidth]{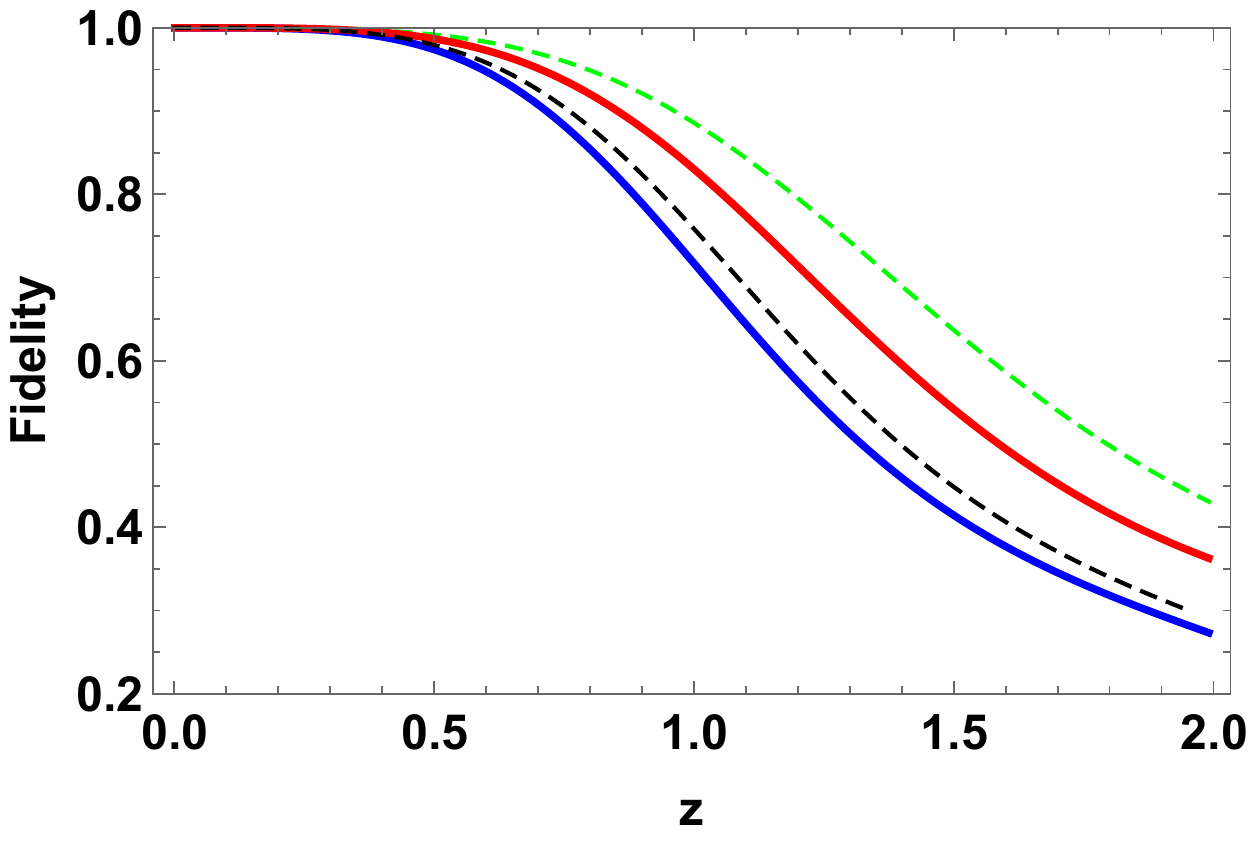}
  \caption{}
  \label{fig1_4}
\end{subfigure}
\label{fig1}
  \caption{Fidelity behavior for both GHA and generalized su(1,1) catlike states for the perturbed harmonic oscillator as function of time, the parameter $|z|$ and different parameters of deformation $a,d$ and $e$. (a) The time evolution of the fidelity for  various catlike states with $|z|=1$ and $d=2e=0.2$. The blue line is for the GHA catlike state with $a=0.5$, red line is for generalized su(1,1) catlike state with $a=0.5$, black line is for  GHA catlike state with $a=0.7$  and green line is for generalized su(1,1) catlike state with $a=0.7$. (b) The time evolution of the fidelity for  various catlike states with $|z|=1$ and $d=2e=0.2$. The blue line is for GHA  catlike state with $a=0.2$, red line is  for generalized su(1,1) catlike state with $a=0.2$,  black line is for  GHA catlike state with $a=0.5$ and green line is for generalized su(1,1) catlike state with $a=0.5$. (c) The time evolution of the fidelity for  various catlike states with $|z|=1$ and $a=0.7$. The blue line is for GHA  catlike state with   $d=2e=0.4$, red line is  for generalized su(1,1) catlike state with $d=2e=0.4$, black line is for  GHA catlike state with  $d=2e=0.2$ and green line is for generalized su(1,1) catlike state with  $d=2e=0.2$. (d)  The fidelity  as a function of $|z|$ for  various catlike states with $t=1$ and  $d=2e=0.2$. The blue line is for GHA catlike state with $a=0.5$ , red line is for  generalized su(1,1) catlike state with $a=0.5$, black line is for  GHA catlike state with $a=0.7$ and green line is for generalized su(1,1) catlike state with $a=0.7$.}
  \label{figure1}
\end{figure}

We now study  the interaction between the perturbed  oscillator presented above with a large environment composed by an infinite collection of harmonic oscillators. According to \cite{Milburn}, the time evolution of the density operator $\hat{\rho}$ of the deformed harmonic oscillator is described by the following master equation
\begin{equation}
\dfrac{d\hat{\rho}(t)}{dt}=\gamma a \hat{\rho}(t) a^\dagger-\dfrac{\gamma}{2}\{a^\dagger a,\hat{\rho}(t)\},
\end{equation}
where $\gamma$ is the damping coefficient.  $a$ and $a^\dagger$ are  the annihilation and creation operators of the harmonic oscillator, respectively.\\
The density operator $\hat{\rho}(t)$ reads as
\begin{equation}\label{sj}
\hat{\rho}(t)=\sum_{j=0}^\infty S_j(t)\hat{\rho}(0)S_j^\dagger(t),
\end{equation}
where
\begin{equation}
S_j(t)=\sum_{n=j}^\infty \sqrt{\dfrac{n!}{(n-j)! j!}}\text{e}^{-(n-j)\gamma t/2}(1-\text{e}^{-\gamma t})^{j/2} \ket{n-j}\bra{n}.
\end{equation}
In the following, we will consider that at $t=0$,  $\hat{\rho}(0)=\ket{\psi_+ }\bra{\psi_+ }$, where $\ket{\psi_+ }$ is given in (\ref{psi2}).
In order to quantify the robustness  of catlike states, we use the fidelity defined by
\begin{equation}\label{fid}
F(t)=\text{Tr}(\hat{\rho}(t)\hat{\rho}(0)).
\end{equation}
This measures how states $\hat{\rho}(t)$ are evolved compared with the initial state $\hat{\rho}(0)$. By using (\ref{sj}), the fidelity (\ref{fid})  of the states  (\ref{psi2}) becomes
\begin{multline}\label{fidt}
F(t)=(\mathcal{N}_+|z|)^4(N(|z|))^4\sum_{j=0}^\infty\sum_{n,m=j}^\infty  \sqrt{\dfrac{n!m!}{(n-j)!(m-j)! (j!)^2}}\text{e}^{-(m+n-2j)\gamma t/2}(1-\text{e}^{-\gamma t})^{j}\\ \times\dfrac{z^n+(-z)^n}{\mathcal{N}_{n-1}!}\dfrac{z^m+(-z)^m}{\mathcal{N}_{m-1}!}\dfrac{z^{n-j}+(-z)^{n-j}}{\mathcal{N}_{n-j-1}!}\dfrac{z^{m-j}+(-z)^{m-j}}{\mathcal{N}_{m-j-1}!}.\hspace{2cm}
\end{multline}
where we have considered that $z$ is a real number rather than complex.
The fidelity of the generalized Heisenberg algebra coherent states superposition known as GHA catlike states can be easily computed by replacing $\mathcal{N}_{n-1}$ by  $N_{n-1}$ in (\ref{fidt}).
In the figures (\ref{fig1_1}), (\ref{fig1_2}) and (\ref{fig1_3}) we show the time evolution of the fidelity of the generalized su(1,1) and GHA catlike states of the perturbed harmonic oscillator having the energy spectrum (\ref{varepsilon}), $\varepsilon_{n}=n+\dfrac{an+e}{n+d}$, for different values of deformation parameters $a$, $d$ and $e$.  In all numerical simulations we have taken the damping rate   $\gamma=1$. We see from these figures that for the same values of the parameters $a$, $d$ and $e$, the generalized su(1,1) catlike states are always more resistant against decoherence than the GHA ones. In addition, we see that perturbed harmonic oscillators having $a=0.7$  are more robust to decoherence   than the ordinary  harmonic oscillator associated with $a=0.5$ and $d=2e$. Furthermore, the figure (\ref{fig1_3}) shows that  the robustness to decoherence of the perturbed oscillators depends also on the parameters $d$ and $e$. Therefore, we can conclude that the catlike states' resistance to  decoherence depends on the corresponding algebra from which they are constructed, i.e, the generalized su(1,1) and the GHA. Varying the different
parameters of deformations $a$, $d$ and $e$, we can find perturbed harmonic oscillators which are more resistant to the decoherence caused by the interaction with the environment compared to the non perturbed harmonic oscillator. The figure (\ref{fig1_4}) shows how the resistance against decoherence of the perturbed harmonic oscillators varies with the amplitude $|z|$. We see that all catlike states ( with different values of the parameter $a$) are more robust against decoherence for very small values of $|z|$. For all catlike states, the fidelity decreases monotonically with the time and tends to zero for long values of time indicating that the final state does not present quantum coherence. Therefore, the final state can not be written as a coherent superposition of quantum states (loss of coherence). However, quantum coherence of perturbed harmonic oscillators can be more or less resistant to decoherence depending on the choice of the deformation parameters.
\section{Physical properties of generalized catlike states under decoherence}\label{sec5}
\subsection{Photon distribution}\label{Photon distribution}
\begin{figure}
\begin{subfigure}{.5\textwidth}
  \centering
  \includegraphics[width=.6\linewidth]{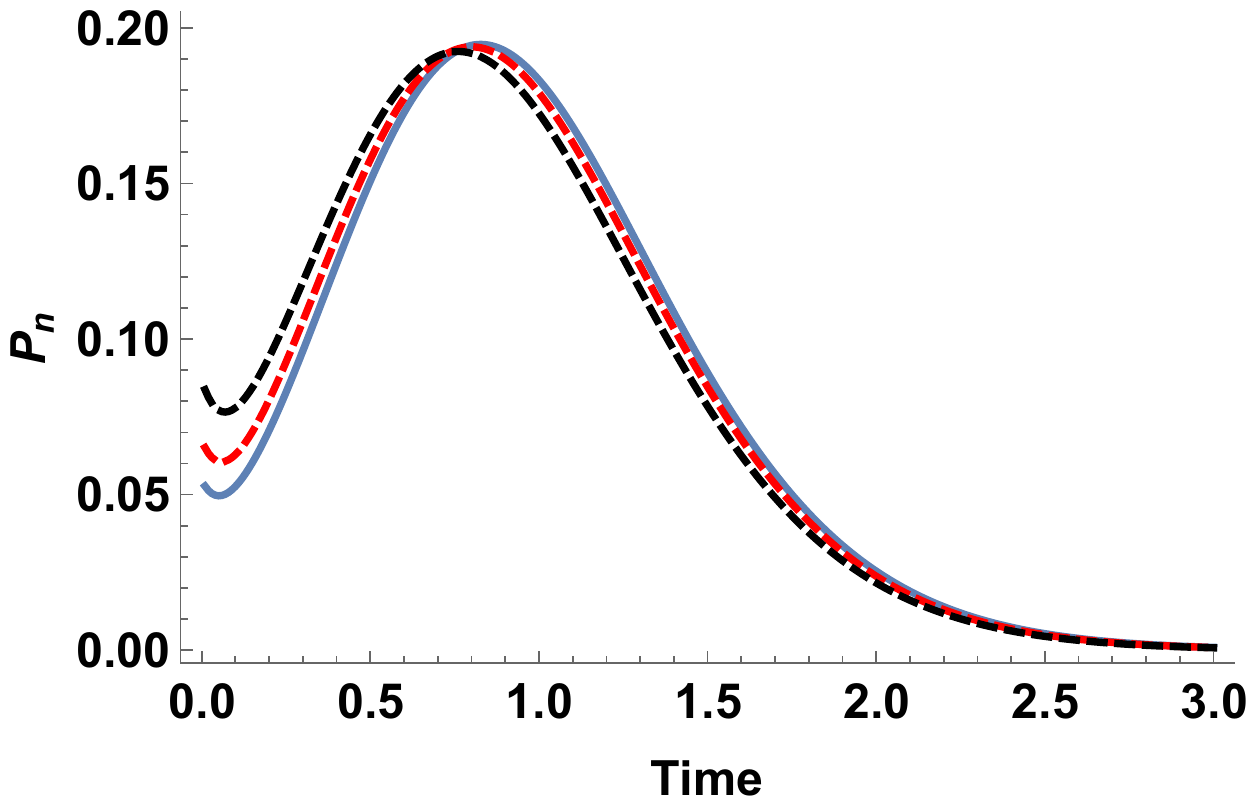}
  \caption{}
  \label{fig2_1}
\end{subfigure}%
\begin{subfigure}{.5\textwidth}
  \centering
  \includegraphics[width=.6\linewidth]{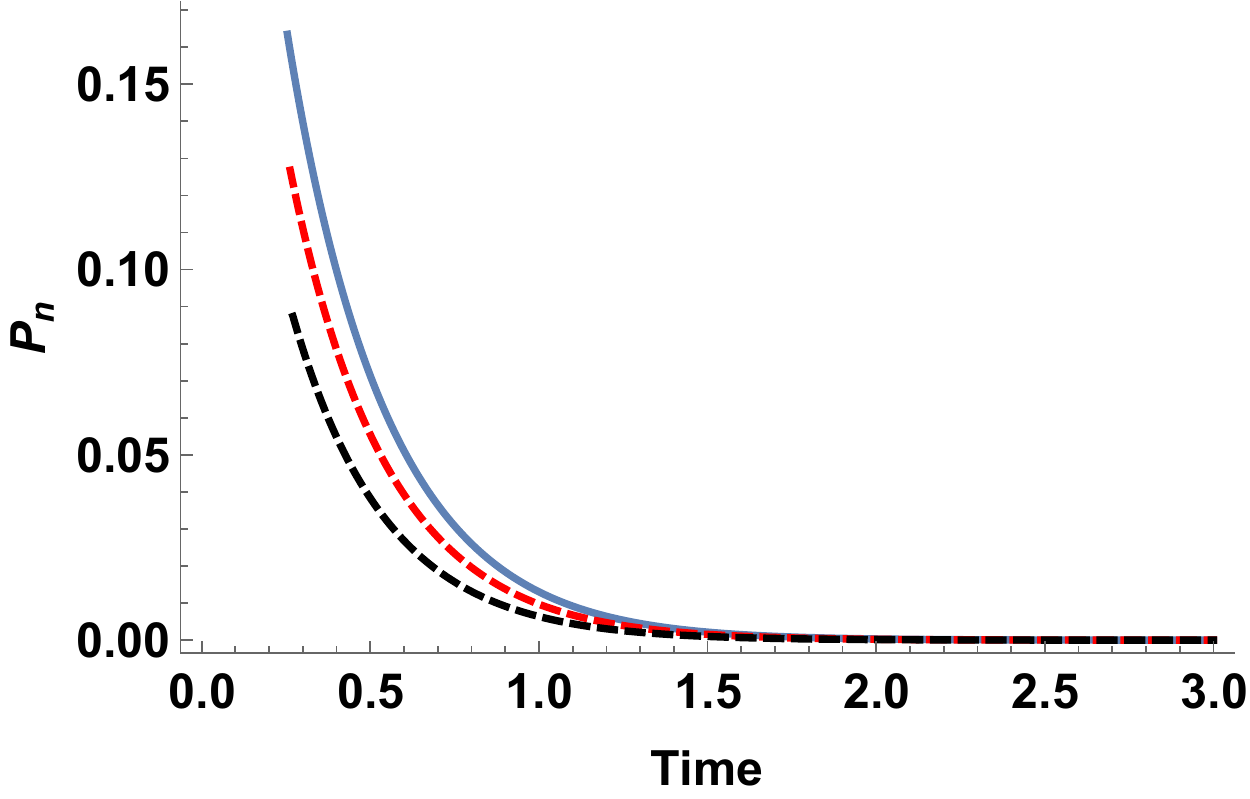}
  \caption{}
  \label{fig2_2}
\end{subfigure}
\\
\begin{subfigure}{.5\textwidth}
  \centering
  \includegraphics[width=.6\linewidth]{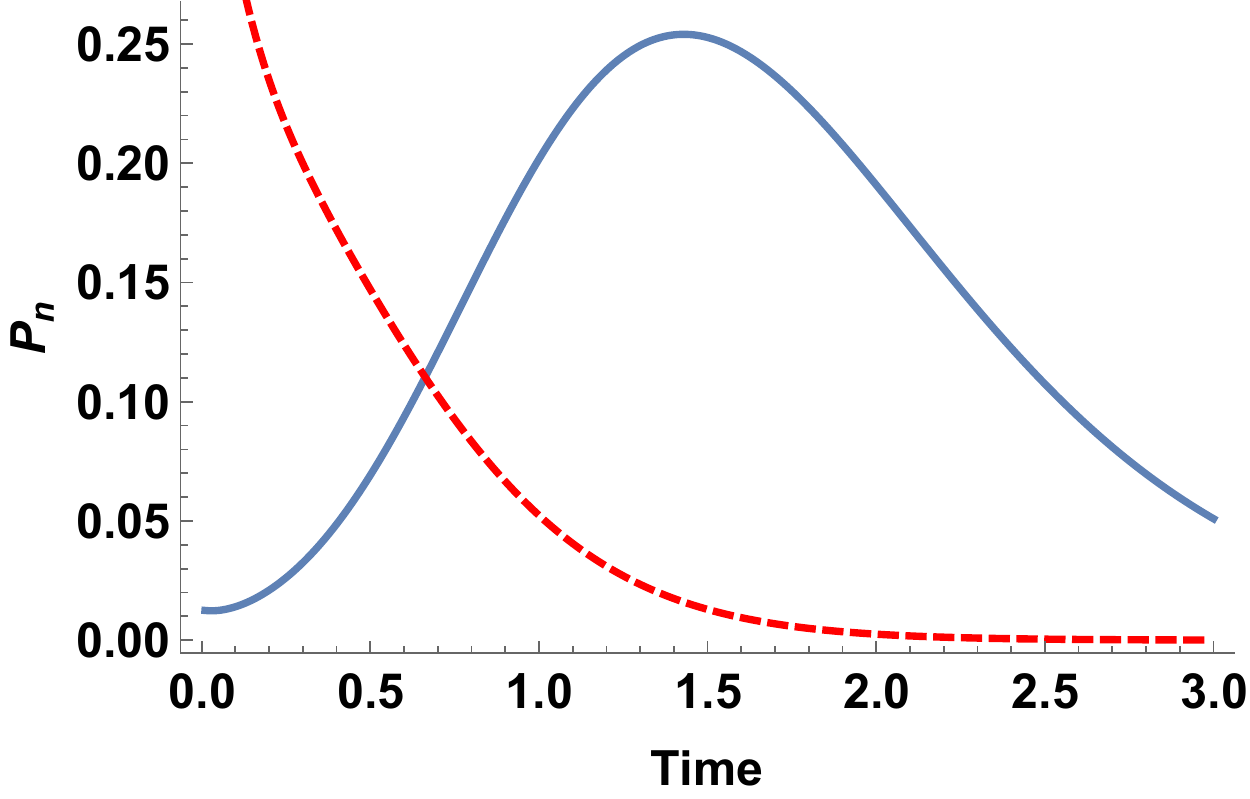}
  \caption{}
  \label{fig2_3}
\end{subfigure}
\begin{subfigure}{.5\textwidth}
  \centering
  \includegraphics[width=.6\linewidth]{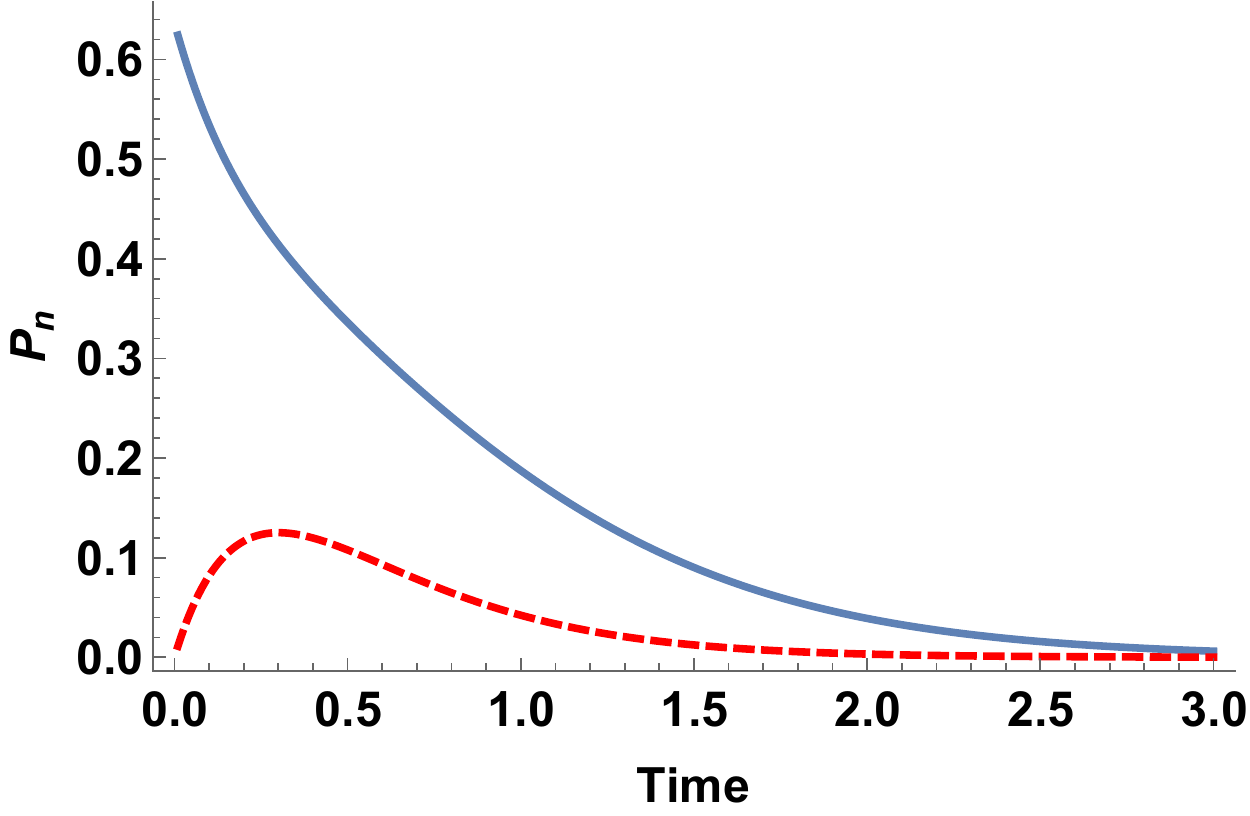}
  \caption{}
  \label{fig2_4}
\end{subfigure}
\begin{subfigure}{.5\textwidth}
	\centering
	\includegraphics[width=.6\linewidth]{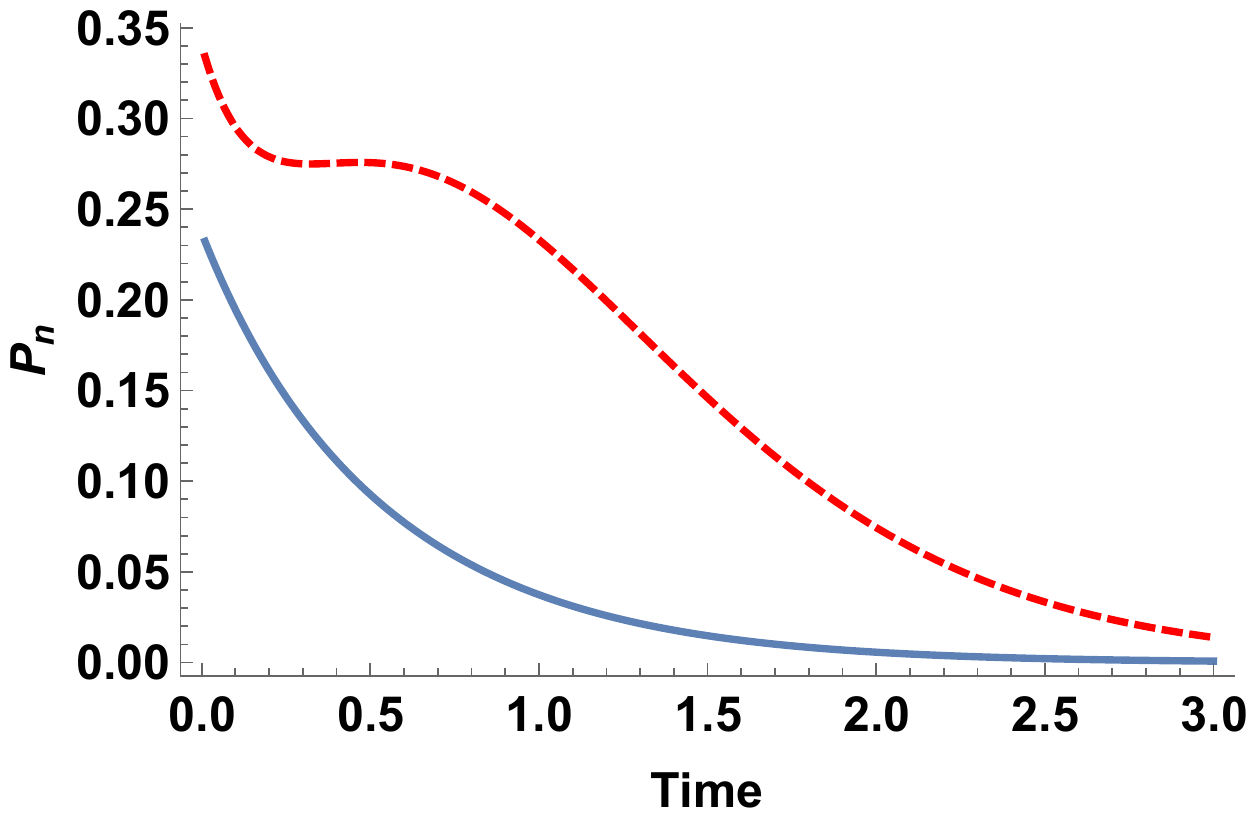}
	\caption{}
	\label{fig2_5}
\end{subfigure}
\begin{subfigure}{.5\textwidth}
	\centering
	\includegraphics[width=.6\linewidth]{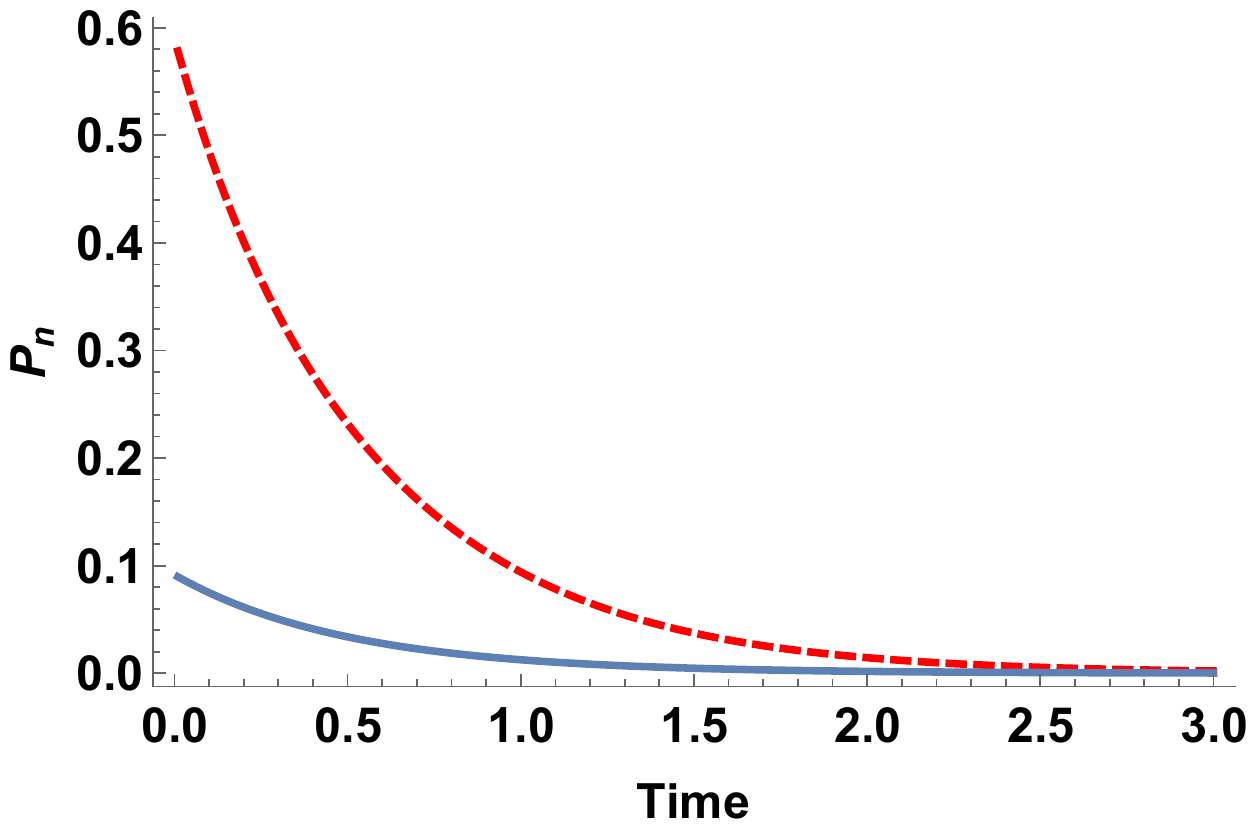}
	\caption{}
	\label{fig2_6}
\end{subfigure}
\label{fig2}
  \caption{The time evolution of the photon distribution function  of  GHA  and generalized su(1,1) catlike states for the perturbed harmonic oscillator  for various physical parameters. (a) The time evolution of the photon distribution function  for GHA catlike states with $|z|=3$, $n=4$ and $d=2e=0.2$. The blue line is for $a=0.2$, red line is for $a=0.5$ and black line is for $a=0.9$. (b) The time evolution of the photon distribution function  for generalized su(1,1) catlike states with $|z|=3$, $n=4$ and $d=2e=0.2$. The blue line is for $a=0.2$, red line is for $a=0.5$ and black line is for $a=0.9$. (c) The time evolution of the photon distribution function  for GHA  catlike states with $|z|=3$, $a=0.7$ and $d=2e=0.2$. The blue line is for $n=2$ and red line is  for $n=3$. (d)  The time evolution of the photon distribution function  for generalized su(1,1)  catlike states with $|z|=3$, $a=0.7$ and $d=2e=0.2$. The blue line is for $n=2$ and red line is  for $n=3$. (e) The time evolution of the photon distribution function  for GHA  catlike states with $n=2$, $a=0.9$ and $d=2e=0.2$. The blue line is for $|z|=1$ and red line is  for $|z|=2$. (f) The time evolution of the photon distribution function  for generalized su(1,1)  catlike states with $n=2$, $a=0.9$ and $d=2e=0.2$. The blue line is for $|z|=1$ and red line is  for $|z|=2$.}
  \label{figure2}
\end{figure}

We now analyze the probability of  finding $n$ photons in generalized catlike states constructed in section (\ref{sec3}). For a density operator $\hat{\rho}(t)$, this probability is defined by
\begin{equation}\label{Pn}
    P_n(t)=\text{Tr}(\hat{\rho}(t)\ket{n}\bra{n}).
\end{equation}
By using the results found in \cite{Halouch,Sanjib}, the photon distribution function of generalized su(1,1) catlike states can be given by
\begin{equation}\label{pro}
    P_n(t)=(N(|z|))^2(\mathcal{N}_+(|z|))^2\sum_{j=0}^\infty \dfrac{(n+j)!}{n!j!}(\text{e}^{-\gamma t})^n(1-\text{e}^{-\gamma t})^j\times\dfrac{(1+(-1)^{n+j})^2|z|^{2n+2j}}{(\mathcal{N}_{n+j-1}!)^2}.
\end{equation}
The photon distribution of GHA catlike states can be obtained by replacing $\mathcal{N}_n$ by $N_n$ in (\ref{pro}). In figure (\ref{figure2}), we show the time evolution of the photon distribution function  for  both GHA and generalized su(1,1) catlike states in terms of different parameters of deformation, the amplitude $|z|$ and the number of photons $n$. Analyzing this figure, we can see that the behavior of the time evolution of the photon distribution of catlike states depends on the corresponding algebraic structure (generalized su(1,1) or GHA).  We also see that for all catlike states, $P_n(t)$ tends to zero as the time becomes very large. Therefore, the decoherence causes the loss of photons of  catlike states which are transferred to the environment. The figures (\ref{fig2_1})-(\ref{fig2_3}) show that for  particular values of $a$, the probability of generalized catlike states is shown to be larger than that of the ordinary harmonic oscillator. From the figures (\ref{fig2_1}), (\ref{fig2_2}), we see that  depending on the deformation parameter $a$, we find perturbed oscillators having large photon distribution function.\\
A close examination of figures (\ref{fig2_1}) and (\ref{fig2_3}) shows  that the probability $P_n(t)$ of some GHA catlike states increases to the maximal value. Then, it decreases to its minimal value as the time becomes significantly large indicating the transfer of photons between the system and the bath. Furthermore, from (\ref{fig2_5})-(\ref{fig2_6}), we find that as the amplitude $|z|$ is large as the time for which $P_n(t)\approx0 $ is maximal for different perturbed catlike states.
\subsection{Statistical properties of generalized catlike states}
\begin{figure}[h]
\begin{subfigure}{.5\textwidth}
  \centering
  \includegraphics[width=.8\linewidth]{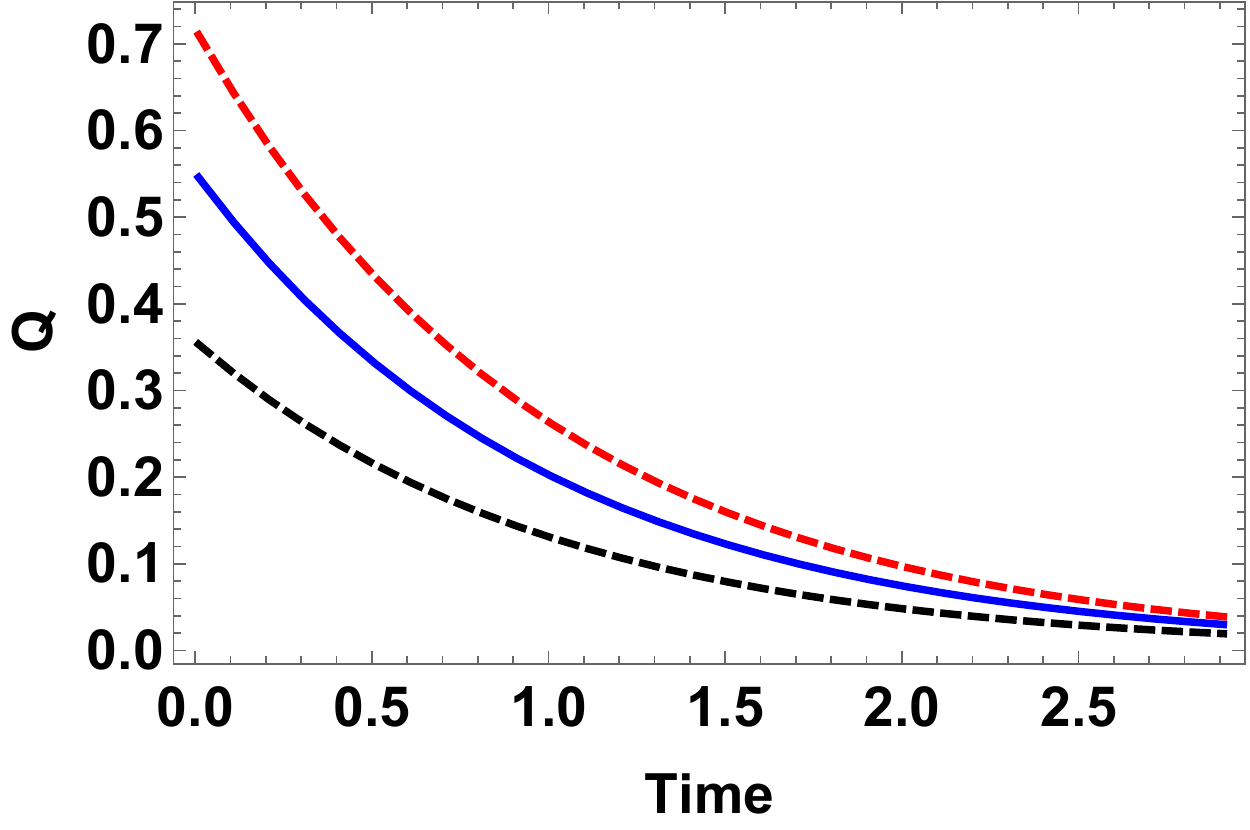}
  \caption{}
\end{subfigure}%
\begin{subfigure}{.5\textwidth}
  \centering
  \includegraphics[width=.8\linewidth]{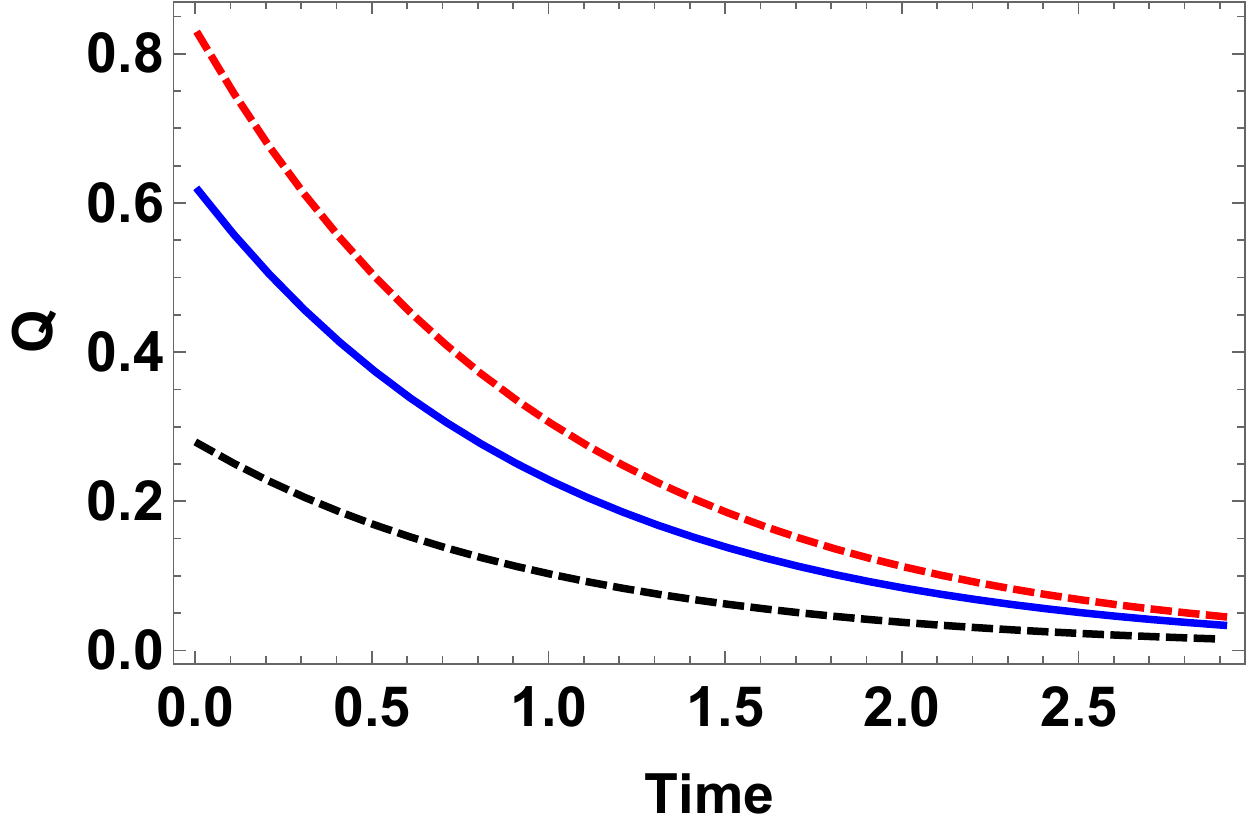}
  \caption{}
\end{subfigure}
  \caption{The time evolution of the Mandel’s parameter for GHA and generalized catlike states for the deformed harmonic oscillator. (a) The time evolution of the Mandel’s parameter  for GHA catlike states with $|z|=1$ and $d=2e=0.2$. The blue line is for $a=0.5$, red line is for $a=0.9$ and black line is for $a=0.2$. (b)  The time evolution of the Mandel’s parameter for generalized su(1,1) catlike states with $|z|=1$ and $d=2e=0.2$. The blue line is for $a=0.5$, red line is for $a=0.9$ and black line is for $a=0.2$.}
  \label{figure3}
\end{figure}
The  statistical  properties of generalized catlike states can be studied by using the Mandel’s parameter
\begin{equation}
    Q=\dfrac{\langle (\Delta \hat{n})^2 \rangle-\langle \hat{n} \rangle}{\langle \hat{n} \rangle},
\end{equation}
where  $\langle \hat{n}\rangle$ is  the  average  number  of  particles  in  the  catlike state in question and $\langle(\Delta \hat{n})^2\rangle=\langle \hat{n}^2 \rangle-\langle \hat{n} \rangle^2$ is its variance. If $Q>0$ $(Q<0)$, we say that the distribution is super-Poissonian (sub-Poissonian) and if $Q=0$, the distribution is Poissonian. To calculate the Mandel’s parameter for generalized catlike states, we use the fact that
\begin{equation}
    \langle \hat{n}(t)\rangle=\sum_{n=0}^\infty n P_n(t)\quad\text{and}\quad  \langle \hat{n}^2(t)\rangle=\sum_{n=0}^\infty n^2P_n(t),
\end{equation}
where $P_n(t)$ is given in (\ref{Pn}). In the figure (\ref{figure3}), we show the time evolution of the Mandel’s parameter for both kinds of catlike states for different values of the deformation parameter $a$ with $|z|=1$. We immediately see  that the distribution of all examined catlike states is super-Poissonian and that $Q$ decreases with increasing values of time and vanishes as the time is significantly large. It is recalled that the distribution of the Glauber coherent state is Poissonian, $Q=0$, and that this coherent state is considered as the most "classical" quantum state for the harmonic oscillator and several measures of nonclassicality of quantum states are related to this coherent state. Then, since for all catlike states, the distribution is super-Poissonian at $t=0$ and decreases  tending to zero for large values of time, we  conclude that the dissipative  interaction with the relevant environment decreases the super-Poissonian distribution of the catlike states of the perturbed harmonic oscillators and the final state has a Poissonian distribution as the Glauber coherent state, i.e, the final state is a classical state. Furthermore, the larger the value of deformation parameter $a$ is, the larger the value of the Mandel’s parameter. This implies that the behavior of the statistical distribution of the deformed catlike states depends on the value of the parameter deformation.
\subsection{Quantum entanglement}
\begin{figure}[h]
\begin{subfigure}{.5\textwidth}
  \centering
  \includegraphics[width=.8\linewidth]{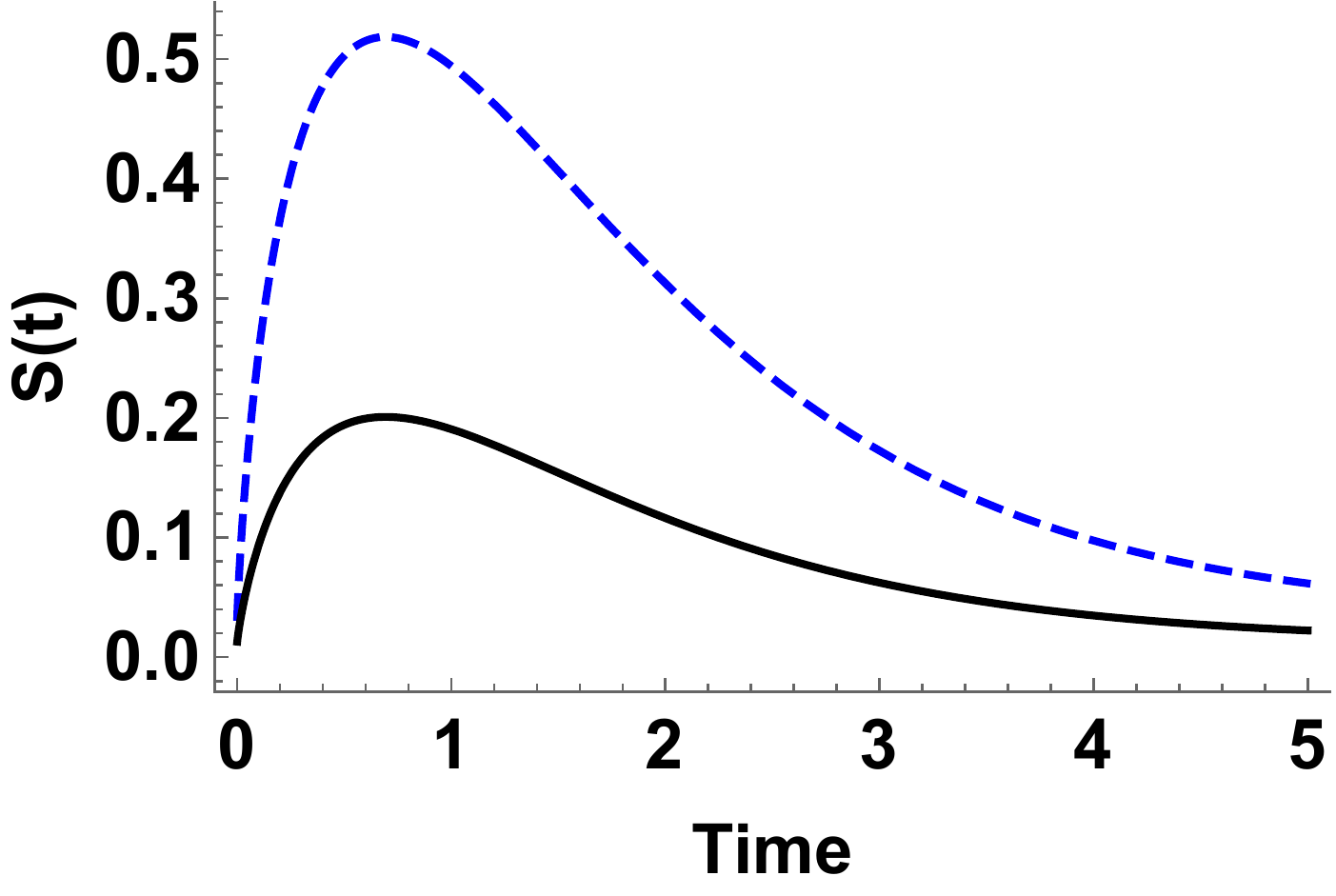}
  \caption{}
\end{subfigure}%
\begin{subfigure}{.5\textwidth}
  \centering
  \includegraphics[width=.8\linewidth]{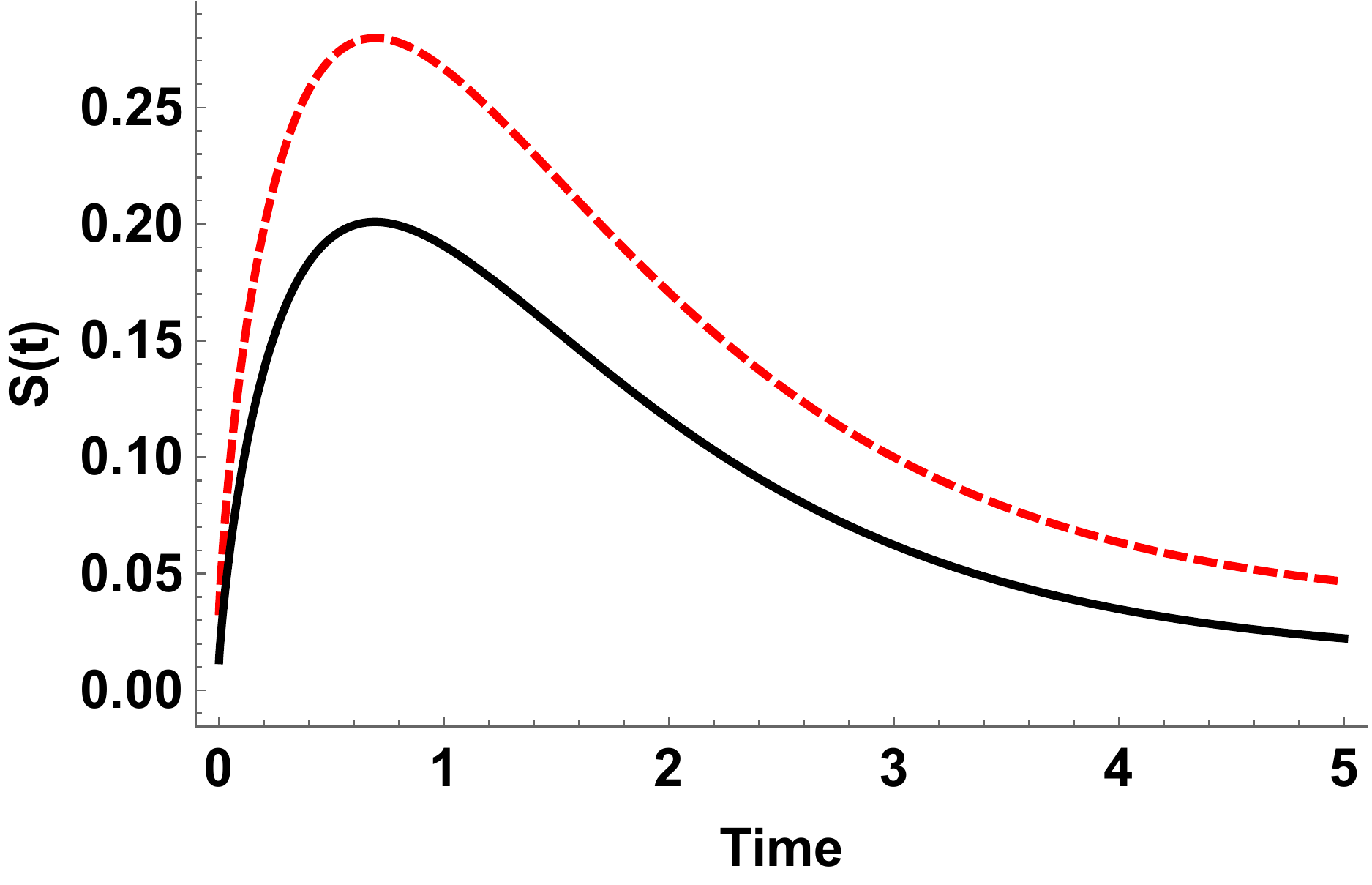}
  \caption{}
\end{subfigure}
  \caption{The time evolution of the von Neumann entropy for generalized su(1,1) catlike states of the perturbed harmonic oscillator in terms of various parameters of deformation and amplitude parameter $|z|$. (a) The time evolution of the von Neumann entropy  for generalized catlike states  with $a=0.9$, $d=0.2$ and $e=0.1$ for two different values of $|z|$. The blue dashed line is for  $|z|=1.5$, black line is for $|z|=1$. (b) The time evolution of the von Neumann entropy  for generalized catlike states  with $|z|=1$, $d=0.2$ and $e=0.1$ for two different values of $a$. The red dashed line is for $a=0.7$, black line is for  $a=0.9$. }
  \label{figure4}
\end{figure}

    Protecting entanglement from decoherence is recognized as a considerable experimental challenge  since  the entanglement is a key resource for quantum technologies. Thence, several authors have studied the effect of decoherence on entanglement of catlike states for different physical systems. Here, we aim to quantify the amount of entanglement of  generalized su(1,1) catlike states constructed in section (\ref{su(1,1) deformed states superposition}) for the deformed harmonic oscillator whose spectrum is given in (\ref{varepsilon}) under decoherence caused by the interaction with a large environment composed by an infinite collection of harmonic oscillators. We will study the effect of deformation parameters and amplitude parameter $|z|$ on the entanglement of generalized su(1,1) catlike states. For this purpose, we appeal to the von Neumann  entropy which is a good quantifier of entanglement. The  von Neumann  entropy of the density operator $\hat{\rho}$ is given by
\begin{equation}
   S(t)=-\text{Tr}(\hat{\rho}(t)\ln{(\hat{\rho}(t))}=-\sum_{i}\lambda_i(t)\ln{(\lambda_i(t))},
\end{equation}
where $\lambda_i$ are the eigenvalues of the reduced density matrix $\hat{\rho}(t)$ of the generalized catlike states. In the figure (\ref{figure4}), we show the behavior of the time evolution of the von Neumann entropy. Immediately, we observe that the entropy increases quickly and reaches its maximum value. Then, it decreases and vanishes for large values of time. Thus, we can conclude that the decoherence  affects on  the correlations between the catlike states  and the  environment. The von Neumann entropy vanishes for long values of time $t$ indicates that the state describing the system when time is significantly large, is pure and not entangled with the environment.  Interestingly, the entanglement degree is shown to be large as the amplitude $|z|$ is large. It depends also on different parameters of
deformation. This is relevant because it gives us the possibility to find cases in which  the entanglement loss is retarded in the time evolution.  It is worth to mention that we have treated the von Neumann entropy for higher values of $|z|$ and the entanglement is shown to be larger as $|z|$ increases. \\

In the section (\ref{Photon distribution}), we have seen that the fidelity vanishes also for large values of the time for all generalized catlike states of the perturbed harmonic oscillator. This indicates that in this range
of time, $t>>1$, the quantum coherence (the interference property) of the state of the
perturbed oscillator are lost due to the interaction with the environment. Furthermore, The Mandel’s parameter $Q$ and the photon distribution function tend to zero as the time is significantly large. This can be explained by the fact that the interaction of the system with the bath transfers all photons from the system to the environment, as $P_n(t)\approx0$ when $t\to\infty$, and leaves the system in a vacuum state which has a Poissonian distribution, i.e, $Q\approx0$ when $t\to \infty$. This state is pure and not entangled.
Precisely, it can not be written as a coherent superposition of other quantum states.
For such a reason, the fidelity and the  von Neumann entropy vanish for large values of time.\\

Consequently, the interaction of the perturbed harmonic oscillators with a collection of harmonic oscillators decreases the fidelity, the amount of their entanglement with the environment, the super-Poissonian distribution and the photon distribution function. However, these properties  can be preserved and retarded by varying the different parameters of deformation.
\section{Conclusion}\label{sec6}
In this paper, we have built the catlike states associated with the generalized su(1,1) algebra for a four-parameter deformed oscillator. In terms of different deformation parameters, we have investigated the resistance of these states under decoherence caused
by a dissipative interaction with an environment modeled by an infinite collection of
harmonic oscillators. Subsequently, we have shown that the generalized su(1,1) catlike states are always more robust under decoherence than the GHA catlike states and that the robustness against the decoherence depends on different parameters of deformations. Among others, we have found that the different parameters of deformations  give the possibility to find cases more resistant than the catlike states of the ordinary harmonic oscillator. It has been revealed that the  resistance  of the catlike states  depends on the algebraic structure from which they are constructed. This may open new windows and perspectives to construct catlike states more robust under decoherence and to the experimental observation of nonclassical features of quantum systems. Moreover, we have studied the time evolution of the photon distribution function, the Mandel’s parameter and the von Neumann entropy for different catlike states and in
terms of different physical parameters. Additionally, we have shown that,  depending on parameters of deformation and amplitude parameter $|z|$, the photon distribution function and the degree of entanglement of the perturbed oscillators can be more preserved in the time evolution.
 \section*{Acknowledgement}
  We would like to acknowledge the partial support  from  ICTP through AF-14.


\begin{thebibliography}{100}
	\bibitem{shrodinger} E. Shcr\"odinger, \emph{Naturwissenschaften}, Vol. 14, 664 (1926).
	\bibitem{Gazeau}  J. P. Gazeau ," Coherent States in Quantum Physics", \emph{Wiley-VCH,Dewey}, (2009).
\bibitem{Zhang} W. Zhang, D. Feng, and R. Gilmore, "Coherent states: Theory and some applications", \emph{Rev. Mod. Phys}, Vol. 62, 867(1990).
	\bibitem{Glauber} R. J. Glauber, "Photon \text{C}orrelations" \emph{Phys. Rev. Lett}, Vol. 10, 84 (1963).
	\bibitem{Klauder1} J. R. Klauder, "Continuous‐Representation Theory. I. Postulates of Continuous‐Representation Theory", \emph{J. Math. Phys}, Vol. 4, 1055 (1963).
	\bibitem{Klauder2} J. R. Klauder, "Continuous‐Representation Theory. II. Generalized Relation between Quantum and Classical Dynamics ", \emph{J. Math. Phys}, Vol. 4, 1058 (1963).
	\bibitem{PhysRevLett.10.277} E. C. G. Sudarshan, "Equivalence of \text{S}emiclassical and \text{Q}uantum \text{M}echanical \text{D}escriptions of \text{S}tatistical \text{L}ight \text{B}eams" \emph{Phys. Rev. Lett}, Vol. 10, 277  (1963).	\bibitem{PhysRevA.54.4560}  R. L. de Matos Filho and   W. Vogel, "Nonlinear coherent states" \emph{Phys. Rev. A}, Vol. 54, 4560  (1996).
	\bibitem{Klauder3}  J. P. Antoine,  J. P. Gazeau, P. Monceau and J. R. Klauder , "Temporally stable coherent states for infinite well and \text{P}\"oschl\text{-}Teller potentials" \emph{J. Math. Phys}, Vol. 42, 2349  (2001).
	\bibitem{Hassouni2} Y. Hassouni, E. M. F. Curado and  M. A. Rego-Monteiro , "Construction of coherent states for physical algebraic systems" \emph{Phys. Rev. A}, Vol. 71, 022104 (2005).

\bibitem{Curado2}  M. A. Rego-Monteiro, E. M. F. Curado and  Ligia M. C. S. Rodrigues, "Time evolution of linear and generalized Heisenberg algebra nonlinear P\"oschl-Teller coherent states" \emph{Phys. Rev. A}, Vol. 96, 052122 (2017).
\bibitem{hydrogenatom}  E. M. F. Curado, M. A. Rego-Monteiro, Ligia M. C. S. Rodrigues and Y. Hassouni, "Coherent states for a degenerate system: The hydrogen atom" \emph{Physica A: Statistical Mechanics and its Applications}, Vol. 371, 16  (2006).
	\bibitem{Perelomov}  A. M. Perelomov, "Coherent states for arbitrary Lie group", \emph{Comm. Math. Phys}, Vol. 26, 222  (1972).
\bibitem{Klau} J. R.  Klauder and B. S. Skagertan , " Coherent States ", \emph{World Scientific,Singapore}, (1985).
\bibitem{Gilmor}  R.  Gilmor, " Geometry of symmetrized states", \emph{Ann. Phys}, Vol. 74,  391 (1972).
\bibitem{Inomata}  A. Inomata, H. Kuratsuji and C. C. Gerry, "Path Integrals and Coherent States of SU(2)and SU(1,1) ",  \emph{Singapore: World  Scientific}, (1992).
\bibitem{Barut}  A. O. Barut and L. Girardello,  " New "coherent"states associated with non-compact groups", \emph{Commun.  math.  Phys}, Vol. 21, 41  (1971).

\bibitem{Shi}  S. Dong "The SU(2) realization for the Morse potential and its coherent states", \emph{
	Canadian Journal of Physics}, Vol. 80, 129  (2002).
\bibitem{Angelova4}  M. Angelova, A. Hertz and  V. Hussin, "Squeezed coherent states and the one-dimensional \text{M}orse quantum system", \emph{J. Phys. A: Math. Theor}, Vol. 45, 244007  (2012).	

\bibitem{Roy}  B. Roy and P. Roy, "Gazeau–Klauder coherent state for the Morse potential and some of its properties", \emph{Phys. Lett. A}, Vol. 296, 187  (2002).	
\bibitem{Maia}  M. Angelova and V. Hussin, "Generalized and Gaussian coherent states for the Morse potential", \emph{J. Phys. A: Math. Theor}, Vol. 41, 304016  (2008).

\bibitem{Daoud}   M. Daoud   and   D. Popov, "Statistical   properties   of Klauder-Perelomov  coherent  states  for  the  Morse  potential", \emph{International  Journal  of  Modern  Physics  B}, Vol. 18, 325  (2004).
\bibitem{PhysRevA.64.013817}  J. R. Klauder,  K. A. Penson and  J. M. Sixdeniers, "Constructing coherent states through solutions of Stieltjes and Hausdorff moment problems", \emph{Phys. Rev. A}, Vol. 64, 013817 (2001).
\bibitem{Madouri} M. El Baz, Y. Hassouni and F. Madouri, "New  construction  of  coherent  states  for  generalized  harmonic oscillators", \emph{Rep. Math. Phys}, Vol. 50, 263 (2002).
\bibitem{Jurco} B. Jurco, "On  coherent  states  for  the  simplest  quantum  groups", \emph{Lett. Math. Phys}, Vol. 21, 51 (1991).
\bibitem{Slusher} R. E. Slusher, L. W. Hollberg, B. Yurke, J. C. Mertz, and J. F. Valley, "Observation of Squeezed States Generated by Four-Wave Mixing in an Optical Cavity", \emph{Phys. Rev. Lett}, Vol.  55, 2409 (1985).
\bibitem{Kimble} H. J. Kimble, M. Dagenais, and L. Mandel, "Photon Antibunching in Resonance Fluorescence", \emph{Phys. Rev. Lett}, Vol. 39, 691 (1977).
\bibitem{Mandel} R. Short and L. Mandel, "Observation of Sub-Poissonian Photon Statistics", \emph{Phys. Rev. Lett}, Vol. 51, 384 (1983).
\bibitem{Yu} T. Yu and J. H. Eberly, "Quantum Open System Theory: Bipartite Aspects", \emph{Phys. Rev. Lett}, Vol. 97, 140403 (2006).
\bibitem{Nielsen} M. A. Nielsen and I. Chuang, "Quantum computation and quantum information", \emph{Cambridge University Press, Cambridge}, (2000).
\bibitem{Horo} R. Horo decki, P. Horo decki, M. Horo decki,and K. Horo decki, "Quantum entanglement", \emph{Reviews of modern physics }, Vol. 81, 865 (2009).

\bibitem{Gardiner}  C. W. Gardiner, "Quantum Noise" \emph{Berlin, Springer}, (1991).
\bibitem{Neumann} J. von Neumann, "Mathematical Foundations of Quantum Mechanics", \emph{Princeton University Press}, (1983).
\bibitem{Nemes} J. I. Kim, M. C. Nemes, A. F. R. de Toledo Piza and H. E. Borges, "Perturbative Expansion for Coherence Loss", \emph{Phys. Rev. Lett}, Vol. 77, 207 (1996).
\bibitem{Zurek}  W. H. Zurek, S. Habib and J. P. Paz, "Coherent states via decoherence", \emph{Phys. Rev. Lett}, Vol. 70, 1187 (1993).
\bibitem{Retamal} J. C. Retamal and N. Zagury, "Stability of quantum states under dissipation", \emph{Phys. Rev. A}, Vol. 63, 032106 (2001).
\bibitem{Kampen} N. G. van Kampen, "A soluble model for quantum mechanical dissipation", \emph{J. Stat. Phys}, Vol. 78, 299 (1995)
\bibitem{Palma} G. M. Palma, K. A. Suominen, A. K. Ekert, "Quantum computers and dissipation", \emph{Proc. Roy. Soc. A}., Vol. 452, 567 (1996).
\bibitem{Leonid} L. Fedichkin, A. Fedorov, V. Privman, "Measures of Decoherence", \emph{arXiv:cond-mat/0303158}
\bibitem{Mancini} S. Mancini and V. I. Manko, "The survival of quantum coherence in deformed-states superposition", \emph{Euro phys. Lett}, Vol. 54, 586 (2001).

\bibitem{Filho} R. L. de Mattos Filho and W. Vogel, "Nonlinear coherent states", \emph{Phys. Rev. A}, Vol. 54, 4560 (1996).
\bibitem{Vogel} Z. Kis, W. Vogel, and L. Davidovich, "Nonlinear coherent states of trapped-atom motion" , \emph{Phys.Rev.A}, Vol. 64, 033401 (2001).
\bibitem{Meekhof} C. Monroe, D. M. Meekhof, B. E. King, and D. J. Wineland, "A “Schrödinger Cat” Superposition State of an Atom", \emph{Science}, Vol. 272, 1131 (1996).
\bibitem{King} C. J. Myatt, B. E. King, Q. A. Turchette, C. A. Sackett,D. Kelpinski, W. M. Itano, C. Monroe, and D. J. Wineland, "Decoherence of quantum superpositions through coupling to engineered reservoirs", \emph{Nature (London)}, Vol. 403, 269 (2000).
\bibitem{2013} E. M. F. Curado, M. A. Rego-Monteiro, and Ligia M. C. S. Rodrigues, "Structure of generalized Heisenberg algebras and quantum decoherence analysis",  \emph{Phys. Rev. A}, Vol. 87, 052120 (2013).
 \bibitem{Twareque} S. Twareque Ali, Lubka Balkova, E. M. F. Curado, J. P Gazeau, M. A. Rego-Monteiro, Ligia M. C. S. Rodrigues, and K. Sekimoto, "Noncommutative reading of the complex plane through Delone sequences", \emph{J. Math. Phys}, Vol. 50, 043517 (2009).
  \bibitem{Monteiro1}  E. M. F. Curado and M. A. Rego-Monteiro, "Thermodynamic properties of a solid exhibiting the energy spectrum given by the logistic map" \emph{Phys. Rev. E}, Vol. 61, 6255 (2000).

   \bibitem{Monteiro2}  E. M. F. Curado and M. A. Rego-Monteiro, "Multi-parametric deformed Heisenberg algebras: a route to complexity" \emph{J. Phys. A: Math. Gen}, Vol. 34, 3253  (2001).

   \bibitem{Hassouni}  E. M. F. Curado, Y. Hassouni  M. A. Rego-Monteiro and Ligia M. C. S. Rodrigues, "Generalized \text{H}eisenberg algebra and algebraic method: The example of an infinite square-well potential" \emph{Phys. Lett. A}, Vol. 372, 3350  (2008).
  \bibitem{Abdessamad1}  Abdessamad  Belfakir and Yassine Hassouni, "Generalized Heisenberg Algebras: periodicity and finite representation", \emph{Physica Scripta}, Vol. 95, 055208 (2020).
  \bibitem{Angelova1}  V. Hussin and  I. Marquette "Generalized \text{H}eisenberg \text{A}lgebras, \text{SUSYQM} and \text{D}egeneracies: \text{I}nfinite \text{W}ell and \text{M}orse \text{P}otential" \emph{SIGMA}, Vol. 7, 024 (2011).
  \bibitem{Bagarello1}  F. Bagarello, E. M. F. Curado and  J. P. Gazeau, "Generalized \text{H}eisenberg algebra and (non linear) pseudo-bosons", \emph{J. Phys. A: Math. Theor}, Vol. 51, 155201  (2018).
  \bibitem{Wybourne} B.G.Wybourne, "Classical Groups for Physicists" \emph{Wiley, New York}, (1974).
\bibitem{Lie Groups} R.Gilmore, "Lie Groups, Lie Algebras and Some of Their Applications", \emph{Dover Publications  } (2012).
  \bibitem{Chakrabarti}  R. Chakrabarti and R. Jagannathan,"A(p, q)-oscillator realization of two-parameter quantum algebras", \emph{J. Phys. A}, Vol. 24, L711(1991).

  \bibitem{Coon} M.  Arik  and  D.  D.  Coon:, "Hilbert  spaces  of  analytic  functions  and  generalized  coherent  states", \emph{J. Math. Phys}, Vol. 17, 524 (1976).
  \bibitem{Perelomov2} A.  M.  Perelomov, "On  the  completeness  of  some  subsystems  of q-deformed  coherent  states", \emph{Helv. Phys.Acta}, Vol. 68 (1996).
\bibitem{Curado00}  E. M. F. Curado and  M. A. Rego-Monteiro, "Hidden symmetries in generalized su(2) algebras", \emph{Physica A}, Vol. 295, 268 (2001).
\bibitem{Curado01}   E. M. F. Curado and  M. A. Rego-Monteiro, "Non-linear generalization of the sl(2) algebra", \emph{Phys. Lett. A}, Vol. 300, 205 (2002).
\bibitem{Milburn} G. J. Milburn, "in Proceedings of the Fifth Summer Schoolon Atomic, Molecular and Optical Physics", \emph{World Scientific, Singapore, }, p. 435 (1999).
\bibitem{Halouch} K. Berrada and H. Eleuch, "Noncommutative deformed cat states under decoherence", \emph{Phys. Rev. D}, Vol. 100, 016020 (2019).
\bibitem{Sanjib} S. Dey, "$q$-deformed noncommutative cat states and their nonclassical properties", \emph{Phys. Rev. D}, Vol. 91, 044024   (2015).

     \end{thebibliography}
\end{document}